%% file: Allerton11.tex
\documentclass[letterpaper, 10pt,twocolumn, conference]{ieeeconf}       

\IEEEoverridecommandlockouts                              
\usepackage{graphics} 
\usepackage{epsfig} 
\usepackage{mathptmx} 
\usepackage{times} 
\usepackage{amsmath} 
\usepackage{amssymb}  
\usepackage{subfigure}
\usepackage{verbatim}
\usepackage{cite}

\newtheorem{theorem}{Theorem}

\usepackage{color}

\newtheorem{corollary}{Corollary}

\newtheorem{definition}{Definition}

\newtheorem{lemma}{Lemma}

{ 
\newtheorem{remark}{Remark}

}
\title{\LARGE \bf Upper Bound on the Capacity of Gaussian Channels with Noisy Feedback}

\author{ \parbox{5 in}{\centering Chong Li and Nicola Elia
         \thanks{This work was supported by NSF under grant number ECS-0901846}\\
         Department of Electrical and Computer Engineering, Iowa State University\\
         Ames, IA, 50011\\
         Email: $\lbrace$chongli, nelia$\rbrace$@iastate.edu\\}
}

\begin{document}

\maketitle \thispagestyle{empty} \pagestyle{empty}
\begin{abstract}
We consider an additive Gaussian channel with additive Gaussian noise feedback. We first derive an upper bound on the n-block capacity (defined by Cover \cite{cover89}). It is shown that this upper bound can be obtained by solving a convex optimization problem. With stationarity assumptions on Gaussian noise processes, we characterize the limit of the n-block upper bound and prove that this limit is the upper bound of the noisy feedback (shannon) capacity.
\end{abstract}


\section{Introduction}
\begin{figure}
\includegraphics[scale=0.3]{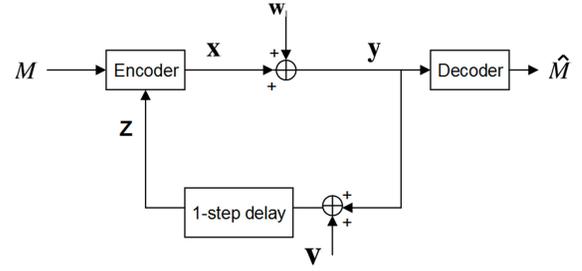}
\caption{Gaussian channels with additive Gaussian noise feedback}
\label{noisyfb_scheme}
\end{figure}

\indent We consider a time-varying additive Gaussian channel with time-varying additive Gaussian feedback. See Fig.\ref{noisyfb_scheme}. $M$ is a message index where $M\in\lbrace 1,2,3,\cdots,2^{nR}\rbrace$. The additive Gaussian channel is modeled as
\begin{equation*}
Y_i=X_i+W_i \qquad i=1,2,\cdots
\end{equation*}
where the gaussian noise $\lbrace W_i \rbrace_{i=0}^{\infty}$ satisfies $W^n\sim \mathbb{N}_n(0,\mathbf{K}_{w,n})$ for all $n\in \mathbf{Z}^+$. Similarly, the additive Gaussian feedback is modeled as
\begin{equation*}
Z_i=Y_i+V_i \qquad i=1,2,\cdots
\end{equation*}
where the gaussian noise $\lbrace V_i \rbrace_{i=0}^{\infty}$ satisfies $V^n\sim \mathbb{N}_n(0,\mathbf{K}_{v,n})$ for all $n\in \mathbf{Z}^+$. Noise $V$ and $W$ are assumed to be independent. Notice that we have not assumed stationarity on $W$ and $V$. The channel input $X_i$ is generated based on $M$ and $Z^{i-1}$, satisfying
\begin{equation*}
\frac{1}{n}\sum_{i=1}^{n}\mathbf{E}X_i^2(M,Z^{i-1})\leq P.
\end{equation*}
Since the capacity of this noisy feedback Gaussian channel is difficult to characterize, we wish to find a tight upper bound on the capacity in this paper.\\
\indent In retrospect, additive Gaussian channels have been studied since the birth of ``Information Theory''. When there is no feedback (i.e. $Z_i=0$ for all $i$), the channel input $X_i$ is independent of the previous channel outputs. The n-block capacity is characterized in \cite{cover89} as
\begin{equation*}
C_n=\underset{\quad tr(\mathbf{K}_{x,n})\leq nP}{\rm max} \quad \frac{1}{2n}\log{\frac{\det{(\mathbf{K}_{w,n}+\mathbf{K}_{x,n})}}{\det{\mathbf{K}_{w,n}}}}
\end{equation*}
where the maximum is taken over all positive semidefinite matrices $\mathbf{K}_{x,n}$. Here, the n-block capacity can be thought of as \textit{the capacity in bits per transmission if the channel is to be used for the time block $\lbrace 1,2,\cdots,n\rbrace$}\cite{cover89}. If we assume the stationarity on the process $\lbrace W_i \rbrace_{i=0}^{\infty}$, it is well-known that the nonfeedback (Shannon) capacity is characterized by water-filling on the noise power spectrum. Specifically,
\begin{equation*}
C=\frac{1}{4\pi}\int_{-\pi}^{\pi}\log\frac{\max\lbrace \mathbb{S}_{w}(e^{i\theta}),\lambda \rbrace}{\mathbb{S}_{w}(e^{i\theta})}d\theta.
\end{equation*}
where $\mathbb{S}_{w}(e^{i\theta})$ is the power spectrum density of the stationary noise process $\lbrace W_i \rbrace_{i=0}^{\infty}$. The water level $\lambda$ should satisfy
\begin{equation*}
\frac{1}{2\pi}\int_{-\pi}^{\pi}\max\lbrace 0, \lambda-\mathbb{S}_{w}(e^{i\theta})\rbrace d\theta = P.
\end{equation*}
Note that the initial idea of water-filling comes from Shannon \cite{shannon49}. When there is a perfect feedback (i.e. $Z_i=Y_i$ for all $i$), the n-block feedback capacity is notably characterized in \cite{cover89} as
\begin{equation*}
C_{fb,n}=\underset{\mathbf{B}_n,\mathbf{K}_{s,n}}{\rm max} \frac{1}{2n}\log{\frac{\det{((\mathbf{I}_n+\mathbf{B}_n)\mathbf{K}_{w,n}(\mathbf{I}_n+\mathbf{B}_n)^T+\mathbf{K}_{s,n})}}{\det{\mathbf{K}_{w,n}}}}
\end{equation*}
where the maximum is taken over all positive semidefinite matrices $\mathbf{K}_{s,n}$ and all strictly lower triangular matrices $\mathbf{B}_{n}$ satisfying
\begin{equation*}
tr(\mathbf{K}_{s,n}+\mathbf{B}_{n}\mathbf{K}_{w,n}\mathbf{B}_{n}^{T})\leq nP.
\end{equation*}

Similar to the nonfeedback case, if we assume the stationarity on the process $\lbrace W_i \rbrace_{i=0}^{\infty}$, the perfect feedback (Shannon) capacity is characterized in \cite{Kim10} as
\begin{equation}
C_{fb}=\sup_{\mathbb{S}_s, \mathbb{B}}\frac{1}{4\pi}\int_{-\pi}^{\pi}\log \frac{\mathbb{S}_s(e^{i\theta})+|1+\mathbb{B}(e^{i\theta})|^2 \mathbb{S}_w(e^{i\theta})}{\mathbb{S}_w(e^{i\theta})}d\theta.
\label{perfect_capacity01}
\end{equation}
with power constraint
\begin{equation}
\frac{1}{2\pi}\int_{-\pi}^{\pi}\mathbb{S}_s(e^{i\theta})+|\mathbb{B}(e^{i\theta})|^2\mathbb{S}_w(e^{i\theta})d\theta\leq P.
\label{perfect_capacity02}
\end{equation}
Here $\mathbb{B}(e^{i\theta})$ represents all possible strictly causal linear filters. When there is an additive Gaussian noise feedback as shown in Fig.\ref{noisyfb_scheme}, no characterization on the capacity $C_{fb}^{noise}$ has been developed yet, to the present author's knowledge. So far, only few papers have addressed this problem or its variations. \cite{Chance10} and \cite{chong11_ISIT} take Cover-Pombra scheme for the noisy feedback case (colored Gaussian noise) and derive the upper and lower bounds on its maximal achievable rate. Other works focus on the additive white Gaussian noise (AWGN) channel with AWGN feedback. For example, \cite{Kim07} derives the upper and lower bounds on the reliability function and shows that, noise in the feedback link renders the noisy feedback communication fundamentally different from the perfect feedback case. \cite{Wyner69},\cite{Martins08} and \cite{chance11} propose specific coding/decoding schemes based on the notable Schalkwijk-Kailath Scheme \cite{Schalkwijk66}. \\
\indent In this paper, we derive upper bounds on the n-block noisy feedback capacity and the noisy feedback (shannon) capacity, respectively. It is shown that the problem of computing the derived upper bound on the n-block capacity can be transformed into a convex form, which can be solved efficiently by standard technical tools.\\
\indent Notations: Uppercase and corresponding lowercase letters $(e.g. Y,Z,y,z)$ denote random variables and realizations, respectively. $x^n$ represents the vector $[x_1,x_2,\cdots,x_n]^T$ and $x^0=\emptyset$. $\mathbf{I}_n$ represents an $n\times n$ identity matrix. $\mathbf{K}_n>0$ ($\mathbf{K}_n\geq 0$) denotes that the $n\times n$ matrix $\mathbf{K}_n$ is positive definite (semi-definite). $log$ denotes the logarithm base $2$ and $0\log0=0$. The expectation operator over $X$ is presented as $\mathbb{E}(X)$.

\section{Preliminaries}
\indent In this section, we review some definitions and Lemmas in information theory.
\begin{definition}\cite{cover06}
The mutual information $I(X;Y)$ between two random variables with joint density $f(x,y)$ is defined as
\begin{equation*}
I(X;Y)=\int f(x,y)\log\frac{f(x,y)}{f(x)f(y)}dxdy
\end{equation*}
\end{definition}

\indent Let $h(X)$ denote the differential entropy of a random variable X. Then it is clear that
\begin{equation*}
I(X;Y)=h(Y)-h(Y|X)
\end{equation*}

\indent We recall a useful Lemma as follows.
\begin{lemma}\cite{cover06}
Let the random vector $X\in \mathbb{R}^{n}$ have zero mean and covariance $\mathbf{K}_{x,n}=\mathbb{E}XX^T$ (i.e. $\mathbf{K}_{ij}=\mathbb{E}X_iX_j$, $1\leq i,j\leq n$). Then
\begin{equation*}
h(X)\leq \frac{1}{2}\log(2\pi e)^n\det\mathbf{K}_{x,n}
\end{equation*}
with equality if and only if $X\sim \mathit{N}(0,\mathbf{K}_{x,n})$.
\label{pre01}
\end{lemma}

\indent Next, we present the definition of \textit{Directed Information} given by Massey \cite{Massey1990}.
\begin{definition}
The directed information from a sequence $X^n$ to a sequence $Y^n$ is defined by
\begin{equation*}
I(X^n\rightarrow Y^n)=\sum_{i=1}^n I(X^i;Y_i|Y^{i-1}).
\end{equation*}
\end{definition}
\indent We would like to remark that Massey's definition of directed information implicitly restricts the time ordering of random variables $(X^n,Y^n)$ as follows
\begin{equation}
X_1, Y_1, X_2, Y_2,\cdots, X_n, Y_n.
\end{equation}
So we refer the interested readers to \cite{Tati09} for the definition of Directed Information for an arbitrary time ordering of random variables. We next define a channel code for communication channels with noisy feedback.
\begin{definition}(\textit{Channel Code})
A $(n,M,\epsilon_n)$ channel code over time horizon n consists of an index set $\lbrace 1,2,3,\cdots, M\rbrace$, an encoding function $e$: $\lbrace 1,2,\cdots,M\rbrace\times \mathcal{Z}^{n-1}\rightarrow\mathcal{X}^n$, a decoding function g:$\mathcal{Y}^n\rightarrow \lbrace 1,2,\cdots,M\rbrace$ and an error probability satisfying
\begin{equation*}
\frac{1}{M}\sum_{w=1}^M p(w\neq g(y^n)|w)\leq \epsilon_n
\end{equation*}
where $\lim_{n\rightarrow\infty}\epsilon_n=0$.
\end{definition}

\indent We finally recall the Schur complement which will play an important role in the paper.
\begin{definition} \cite{ConvexOpti}
Consider an $n\times n$ symmetric matrix $X$ partitioned as
\begin{equation*}
X=\begin{bmatrix}A & B \\ B^T & C\end{bmatrix}.
\end{equation*}
If $\det{A}\neq 0$, the matrix
\begin{equation*}
S=C-B^TA^{-1}B
\end{equation*}
is called the Schur complement of $A$ in $X$.
\end{definition}
We present some properties of the Schur complement as follows.
\begin{enumerate}
\item $\det X=\det A\det S$.
\item $X\geq 0$ if and only if $A > 0$ and $S > 0$.
\item If $A> 0$, then $X\geq  0$ if and only if $S\geq 0$.
\end{enumerate}

\section{An Upper Bound on the N-block Capacity}
\indent In this section, we first derive an upper bound on the n-block noisy feedback capacity (Theorem \ref{thm03_01}). Without loss of generality, we characterize this upper bound as an optimization problem by adopting the Cover-Pombra scheme (Theorem \ref{thm03_02}). We then transform the optimization problem into a convex form (Corollary \ref{coro_03_01}). The n-block noisy feedback capacity is defined as follows \cite{cover89}.
\begin{definition}
\begin{equation}
C_{fb,n}^{noise}=\max_{\frac{1}{n}tr(\mathbf{K}_{X,n})\leq P}\frac{1}{n}I(M;Y^n).
\end{equation}
\end{definition}

\indent We now define a new quantity and then prove that it is an upper bound of the above n-block capacity.
\begin{definition}
\begin{equation}
\bar{C}_{fb,n}^{noise}=\max_{\frac{1}{n}tr(\mathbf{K}_{X,n})\leq P}\frac{1}{n}I(X^n\rightarrow Y^n|V^n).
\end{equation}
\end{definition}

\begin{theorem}
For a given power constraint $P$,
\begin{equation}
C_{fb,n}^{noise}\leq \bar{C}_{fb,n}^{noise}.
\end{equation}
\label{thm03_01}
\end{theorem}

\begin{proof}
\begin{equation*}
\begin{split}
&I(M;Y^n)\\
=&h(M)-h(M|Y^n)\\
=&h(M)-h(M|Y^n,V^n)-(h(M|Y^n)-h(M|Y^n,V^n))\\
\stackrel{(a)}{=}&h(M|V^n)-h(M|Y^n,V^n)-I(M;V^n|Y^n)\\
=&I(M;Y^n|V^n)-I(M;V^n|Y^n)\\
=&h(Y^n|V^n)-h(Y^n|M,V^n)-I(M;V^n|Y^n)\\
=&\sum_{i=1}^n h(Y_i|Y^{i-1},V^n)-h(Y_i|Y^{i-1},M,V^n)-I(M;V^n|Y^n)\\
\stackrel{(b)}{=}&\sum_{i=1}^n h(Y_i|Y^{i-1},V^n)-h(Y_i|Y^{i-1},M,V^n,X^i)-I(M;V^n|Y^n) \\
\stackrel{(c)}{=}&\sum_{i=1}^n h(Y_i|Y^{i-1},V^n)-h(Y_i|Y^{i-1},X^i, V^n)-I(M;V^n|Y^n)\\
=&\sum_{i=1}^n I(X^i;Y_i|Y^{i-1},V^n)-I(M;V^n|Y^n)\\
=&I(X^n\rightarrow Y^n|V^n)-I(M;V^n|Y^n)\\
\end{split}
\end{equation*}
(a) follows from the fact that $M$ and $V^n$ are independent. (b) follows from the fact that $X^i$ can be determined by $M$ and the outputs of the feedback link (i.e. $Y^{i-1}+V^{i-1}$). (c) follows from the Markov chain $M-(Y^{i-1},X^i,V^n) - Y_i$.\\
\indent Since the conditional mutual information $I(M;V^n|Y^n)\geq 0$, we have
\begin{equation*}
I(M;Y^n)\leq I(X^n\rightarrow Y^n|V^n)
\end{equation*}
\indent The proof is completed.
\end{proof}

\indent Next, we characterize the above upper bound. First of all, we consider a scheme with linear encoding of the feedback signal and Gaussian signaling of the message (Cover-Pombra scheme) as shown in a vector form in Fig.\ref{CP_scheme}. \\

\indent \textit{The channel input signal:}  $X^n=S^n+\mathbf{B}_n(W^n+V^n)$\\
\indent \textit{The channel output signal:} $Y^n=S^n+\mathbf{B}_n(W^n+V^n)+W^n$\\
\indent \textit{The power constraint:} $tr(\mathbf{K}_{s,n}+\mathbf{B}_n(\mathbf{K}_{w,n}+\mathbf{K}_{v,n})\mathbf{B}_n^T)\leq nP$\\
\vspace{1mm}
where $S^n\sim \mathit{N}(0,\mathbf{K}_{s,n})$ is the message information vector and $\mathbf{B}_n$ is an $n\times n$ strictly lower triangular linear encoding matrix. Note that the one-step delay in the feedback link is captured by the structure of matrix $\mathbf{B}_n$. Random variables $S^n$,$V^n$,$W^n$ are automatically assumed to be independent.\\
\indent In the following, we prove that $\bar{C}_{fb,n}^{noise}$ can be characterized by the above coding scheme without losing the optimality.
\begin{theorem}
$\bar{C}_{fb,n}^{noise}$ can be obtained as the optimal objective value of the following optimization problem.
\begin{equation}
\begin{split}
\quad \underset{\mathbf{B}_n,\mathbf{K}_{s,n}}{\rm maximize} &\quad \frac{1}{2n}\log{\frac{\det{((\mathbf{I}_n+\mathbf{B}_n)\mathbf{K}_{w,n}(\mathbf{I}_n+\mathbf{B}_n)^T+\mathbf{K}_{s,n})}}{\det{\mathbf{K}_{w,n}}}}\\
\text{subject to} &\quad tr(\mathbf{K}_{s,n}+\mathbf{B}_n(\mathbf{K}_{v,n}+\mathbf{K}_{w,n})\mathbf{B}_n^T)\leq nP\\
&\quad \mathbf{K}_{s,n}\geq 0 \quad \text{$\mathbf{B}_n$ is strictly lower triangular}\\
\end{split}
\label{n-block upperbound}
\end{equation}
\label{thm03_02}
\end{theorem}

\begin{proof}
First of all, we have
\begin{equation*}
\begin{split}
&I(X^n\rightarrow Y^n|V^n)\\
=&\sum_{i=1}^n I(X^i,Y_i|Y^{i-1},V^n) \\
=&\sum_{i=1}^n h(Y_i|Y^{i-1},V^n)-h(Y_i|Y^{i-1},X^i,V^n)  \\
\stackrel{(a)}{=}&\sum_{i=1}^n h(Y_i|Y^{i-1},V^n)-h(Y_i|Y^{i-1},X^i)  \\
=&\sum_{i=1}^n h(Y_i|Y^{i-1},V^n)-h(X_i+W_i|Y^{i-1},X^i,W^{i-1})\\
=&\sum_{i=1}^n h(Y_i|Y^{i-1},V^n)-h(W_i|W^{i-1})\\
=&h(Y^n|V^n)-h(W^n)\\
\end{split}
\end{equation*}
where (a) follows from the Markov chain $V^n-(Y^{i-1},X^i)-Y_i$.
\begin{figure}
\includegraphics[scale=0.4]{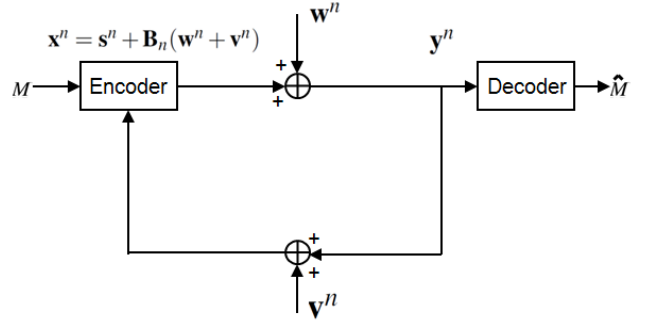}
\caption{Gaussian channels with additive Gaussian noise feedback(Gaussian signalling and linear feedback)}
\label{CP_scheme}
\end{figure}

\indent We next show that maximizing $h(Y^n|V^n)-h(W^n)$ over coding scheme as shown in Fig.\ref{CP_scheme} does not lose the optimality. Since we can not affect the noise entropy (i.e. $h(W^n)$), we need to maximize $h(Y^n|V^n)$ over all possible channel inputs $\lbrace X_i \rbrace_{i=1}^n$. To begin with, we present two insightful observations on the channel inputs $X^n$:
\begin{enumerate}
\item $Y^n|V^n$ should be Gaussian distribution for maximizing $h(Y^n|V^n)$. Since $W^n$ is Gaussian and $Y^n=X^n+W^n$, $X^n|V^n$ must be Gaussian.
\item $X^n$ depends on $W^n+V^n$ instead of $W^n$ and $V^n$ separately since the channel outputs are fed back to the encoder without any encoding.
\end{enumerate}
\indent Therefore, the most general normal causal dependence of $X^n$ on $Y^n$ satisfying the above two observations is the form of $X^n=S^n+\mathbf{B}_n(W^n+V^n)$. Then we have
\begin{equation*}
\begin{split}
&h(Y^n|V^n)-h(W^n)\\
=&h(S^n+\mathbf{B}_n(W^n+V^n)+W^n|V^n)-h(W^n)\\
=&h(S^n+(\mathbf{B}_n+\mathbf{I}_n)W^n|V^n)-h(W^n)\\
=&h(S^n+(\mathbf{B}_n+\mathbf{I}_n)W^n)-h(W^n)\\
\end{split}
\end{equation*}
By Lemma \ref{pre01}, the proof is complete.
\end{proof}

\begin{remark}
The key idea of this theorem is to show
\begin{equation*}
\begin{split}
\bar{C}_{fb,n}^{noise}=&\underset{\text{all coding schemes}}{\rm maximize} \frac{1}{n}I(X^n\rightarrow Y^n|V^n)\\
=&\underset{\text{Cover-Pombra scheme}}{\rm maximize} \frac{1}{n}I(X^n\rightarrow Y^n|V^n)\\
\end{split}
\end{equation*}
We would like to remark that the Cover-Pombra scheme may not be the optimal (capacity-achieving) coding scheme in the noisy feedback case. We herein adopt this coding scheme because it can nicely characterize the proposed upper bound. Definitely, the Cover-Pombra scheme may not apply if we look at a different upper bound.
\end{remark}

\indent The following corollary shows that the optimization problem (\ref{n-block upperbound}) can be transformed into a convex form. This result has been shown in \cite{chong11_ISIT}. For reader's convenience, we give the proof again in the Appendix.

\begin{corollary}
$\bar{C}_{fb,n}^{noise}$ can be obtained as the optimal objective value of the following convex optimization problem.
\begin{equation*}
\begin{split}
\quad \underset{\mathbf{H}_n,\mathbf{B}_{n}}{\rm maximize} &\quad \frac{1}{2n}\log\det \begin{bmatrix} \mathbf{K}_{v,n}^{-1} & \mathbf{B}_n^T \\ \mathbf{B}_n & \mathbf{H}_n \end{bmatrix}-\frac{1}{2n}\log\det(\mathbf{K}_{v,n}^{-1}\mathbf{K}_{w,n})\\
\text{subject to} &\quad tr(\mathbf{H}_n-\mathbf{K}_{w,n}\mathbf{B}_n^T-\mathbf{B}_n\mathbf{K}_{w,n}-\mathbf{K}_{w,n})\leq nP \\
&\quad \begin{bmatrix}  \mathbf{H}_n & \mathbf{I}_n+\mathbf{B}_n^T & \mathbf{B}_n^T\\ \mathbf{I}_n+\mathbf{B}_n & \mathbf{K}_{w,n}^{-1}& \mathbf{0}_{n}\\ \mathbf{B}_n & \mathbf{0}_{n}& \mathbf{K}_{v,n}^{-1} \end{bmatrix}\geq 0\\
&\quad \text{$\mathbf{B}_n$ is strictly lower triangular}\\
\end{split}
\end{equation*}
\label{coro_03_01}
\end{corollary}

\section{An Upper Bound on the Capacity}
\indent As we have shown, the upper bound of the n-block capacity is numerically solvable due to its convex form (interior-point method). As for the Shannon (infinite-block) capacity, however, there is still much work to be done. The problem is that the formula (\ref{n-block upperbound}) may not have a limit as $n\rightarrow \infty$, due to the time-varying nature of the noises $\lbrace W_i \rbrace$ and  $\lbrace V_i \rbrace$. This is the main difficulty for us to develop an upper bound for the Shannon capacity. We herein handle this problem by assuming stationarity on noise $\lbrace W_i \rbrace$ and  $\lbrace V_i \rbrace$. In this section, we first show that under the stationarity assumption the limit of formula (\ref{n-block upperbound}) exists and we develop its limit characterization. Then we prove that the limit characterization is the upper bound of the noisy feedback (Shannon) capacity.\\
\begin{theorem}
Assume that $\lbrace W_i \rbrace$ and  $\lbrace V_i \rbrace$ are stationary processes. Then the limit of formula (\ref{n-block upperbound}) exists and can be characterized as
\begin{equation}
\bar{C}_{fb}^{noise}=\sup_{\mathbb{S}_s, \mathbb{B}}\frac{1}{4\pi}\int_{-\pi}^{\pi}\log \frac{\mathbb{S}_s(e^{i\theta})+|1+\mathbb{B}(e^{i\theta})|^2 \mathbb{S}_w(e^{i\theta})}{\mathbb{S}_w(e^{i\theta})}d\theta
\label{capacity_upperbound}
\end{equation}
with power constraint
\begin{equation}
\frac{1}{2\pi}\int_{-\pi}^{\pi}\mathbb{S}_s(e^{i\theta})+|\mathbb{B}(e^{i\theta})|^2(\mathbb{S}_w(e^{i\theta})+\mathbb{S}_v(e^{i\theta}))d\theta\leq P.
\end{equation}
Here, $\mathbb{S}_s(e^{i\theta})$, $\mathbb{S}_w(e^{i\theta})$ and $\mathbb{S}_v(e^{i\theta})$ are the power spectral density of $\lbrace S_i \rbrace$, $\lbrace W_i \rbrace$ and $\lbrace V_i \rbrace$ respectively. $\mathbb{B}(e^{i\theta})=\sum_{k=1}^{\infty}b_k e^{ik\theta}$ is a strictly causal linear filter.
\label{thm_04_01}
\end{theorem}

\indent The main idea of the proof is taken from \cite{Kim10}. We refer interested readers to the Appendix for the detail. Next, we show that the above limit characterization is the upper bound of the noisy feedback (Shannon) capacity.
\begin{theorem}
Assume that $\lbrace W_i \rbrace$ and  $\lbrace V_i \rbrace$ are stationary processes. Then $C_{fb}^{noise}\leq \bar{C}_{fb}^{noise}$.
\label{thm04_02}
\end{theorem}
\begin{proof}
\begin{equation*}
\begin{split}
&C_{fb}^{noise}\\
\leq &\limsup_{n\rightarrow\infty}\quad \max_{\lbrace X_i\rbrace_{i=0}^n}\frac{1}{n}I(M;Y^n)\\
\stackrel{(a)}{\leq} &\limsup_{n\rightarrow\infty}\quad \max_{\lbrace X_i\rbrace_{i=0}^n}\frac{1}{n}I(X^n\rightarrow Y^n|V^n)\\
\stackrel{(b)}{=} &\limsup_{n\rightarrow\infty}\underset{\mathbf{B}_n,\mathbf{K}_{s,n}}{\rm max} \frac{1}{2n}\log{\frac{\det{((\mathbf{I}_n+\mathbf{B}_n)\mathbf{K}_{w,n}(\mathbf{I}_n+\mathbf{B}_n)^T+\mathbf{K}_{s,n})}}{\det{\mathbf{K}_{w,n}}}}\\
\stackrel{(c)}{=}&\bar{C}_{fb}^{noise}\\
\end{split}
\end{equation*}
where (a) follows from Theorem \ref{thm03_01}, (b) follows from Theorem \ref{thm03_02} and (c) follows from Theorem \ref{thm_04_01}.
\end{proof}

\begin{remark}
Compared with the perfect feedback capacity characterization (\ref{perfect_capacity01}) and (\ref{perfect_capacity02}), feedback noise $\mathbb{S}_v(e^{i\theta})$ only affects the power allocation. If the noise in the feedback link increases (i.e. $\mathbb{S}_v(e^{i\theta})$ grows large in some sense), the feedback benefit in increasing reliable transmission rate vanishes. That is, the noisy feedback system behaves like a nonfeedback system since, due to the power constraint, $\mathbb{B}(e^{i\theta})$ approaches $0$ as $\mathbb{S}_v(e^{i\theta})$ grows.
\end{remark}

\section{Simulation Results}
\begin{figure}
\includegraphics[scale=0.55]{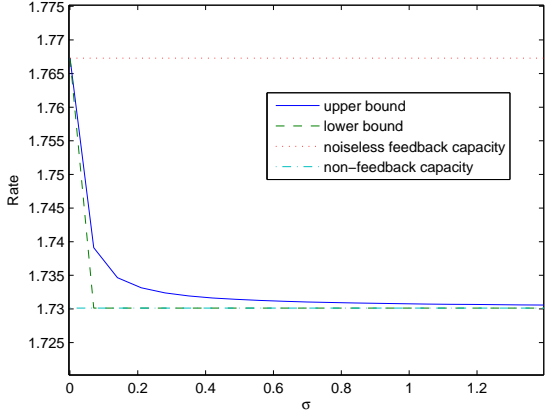}
\caption{The bounds on $C_{fb,n}^{noise}$ of the $1$st-MV channel with $\alpha=0.1$.}
\label{sim_fig01_0.1}
\end{figure}

\indent In this section, we show some simulation results to gain insight on the capacity of Gaussian channels with noisy feedback. The simulation results herein are taken from \cite{chong11_ISIT}. For reader's convenience, we re-present some simulation results here and give a brief discussion. We refer the interested readers to \cite{chong11_ISIT} for more simulation results. We assume that the forward channel is created by a first order moving average ($1$st-MV) Gaussian process. That is,
\begin{equation*}
W_i=U_i+\alpha U_{i-1}
\end{equation*}
where $U_i$ is a white Gaussian process with zero mean and unit variance. We also assume that the feedback link is created by an additive white Gaussian noise with $\mathbf{K}_{v,n}=\sigma \mathbf{I}_n$ ($\sigma\geq 0$). Due to the practical computation limit, we take coding block length $n=30$ and power limit $P=10$. We computed the upper bound of n-block capacity derived in our paper and the lower bound (Theorem $2$ in \cite{chong11_ISIT}) for averaging statistic $\alpha=0.1$ in the $1$st-MV channel, as shown in Fig. \ref{sim_fig01_0.1}. Generally, the plots show that the n-block capacity, which is in the region between the upper and lower bounds, sharply decreases as $\sigma$ grows. When $\sigma$ grows large enough (e.g. $\sigma=0.8$ in Fig.\ref{sim_fig01_0.1}), the feedback rate-increasing enhancement almost shuts off and, thus, the feedback system behaves like a nonfeedback system. Based on this observation, we may claim that the n-block capacity of the Gaussian channel with noisy feedback is sensitive to the feedback noise.\\

\section{Conclusion}
\indent We have derived an upper bound on the n-block capacity of additive Gaussian channels with additive Gaussian noise feedback. As it is shown, this n-block upper bound can be obtained by solving a convex optimization problem. By assuming stationarity on the Gaussian noises, we have characterized the limit of the n-block upper bound, which is the upper bound of the noisy feedback (shannon) capacity. \\
\indent In \cite{chong_jour_nf}, the authors showed that for strong converse finite-alphabet channels with noisy feedback, the capacity is characterized by
\begin{equation*}
C_{FB}^{noise}=\sup_{X}\lim_{n\rightarrow \infty}\frac{1}{n}I(X^n\rightarrow Y^n|V^n)
\end{equation*}
Therefore, we conjecture that the upper bound characterized in this paper should be the true capacity. However, how to prove the achievability of this upper bound remains to be seen.

\section{Appendix}
\input{appendix}

\bibliographystyle{IEEEtran}
\bibliography{ref}

\end{document}

%% file: appendix.tex
\textit{A. The proof of Corollary \ref{coro_03_01}}\\
\begin{proof}
\indent Let $\mathbf{H}_n=(\mathbf{I}_n+\mathbf{B}_n)\mathbf{K}_{w,n}(\mathbf{I}_n+\mathbf{B}_n)^T+\mathbf{K}_{s,n}+\mathbf{B}_n\mathbf{K}_{v,n}\mathbf{B}_n^T$, we have
\begin{equation*}
\begin{split}
&\frac{1}{2}\log{\frac{\det{((\mathbf{I}_n+\mathbf{B}_n)\mathbf{K}_{w,n}(\mathbf{I}_n+\mathbf{B}_n)^T+\mathbf{K}_{s,n})}}{\det{\mathbf{K}_{w,n}}}}\\
=&\frac{1}{2}\log{\frac{\det{(\mathbf{H}_n-\mathbf{B}_n\mathbf{K}_{v,n}\mathbf{B}_n^T)}}{\det{\mathbf{K}_{w,n}}}}.\\
\end{split}
\end{equation*}
\indent We also have
\begin{equation*}
\begin{split}
tr(\mathbf{K}_{x,n})\leq nP &\Leftrightarrow tr(\mathbf{K}_{s,n}+\mathbf{B}_n(\mathbf{K}_{v,n}+\mathbf{K}_{w,n})\mathbf{B}_n^T)\leq nP \\
&\Leftrightarrow tr(\mathbf{H}_n-\mathbf{K}_{w,n}\mathbf{B}_n^T-\mathbf{B}_n\mathbf{K}_{w,n}-\mathbf{K}_{w,n})\leq nP.\\
\end{split}
\end{equation*}

\indent Next, we have the following equivalences by applying the Schur complement. \\
(1).$\det\begin{bmatrix} \mathbf{K}_{v,n}^{-1} & \mathbf{B}_n^T \\ \mathbf{B}_n & \mathbf{H}_n \end{bmatrix}=\det(\mathbf{H}_n-\mathbf{B}_n\mathbf{K}_{v,n}\mathbf{B}_n^T)\det\mathbf{K}_{v,n}^{-1}$.\\
(2).\begin{equation*}
\begin{split}
\mathbf{K}_{s,n}\geq 0 &\Leftrightarrow \mathbf{H}_n-(\mathbf{I}_n+\mathbf{B}_n)\mathbf{K}_{w,n}(\mathbf{I}_n+\mathbf{B}_n)^T-\mathbf{B}_n\mathbf{K}_{v,n}\mathbf{B}_n^T\geq 0\\
&\Leftrightarrow \begin{bmatrix}  \mathbf{H}_n & \mathbf{I}_n+\mathbf{B}_n^T & \mathbf{B}_n^T\\ \mathbf{I}_n+\mathbf{B}_n & \mathbf{K}_{w,n}^{-1}& \mathbf{0}_{n}\\ \mathbf{B}_n & \mathbf{0}_{n}& \mathbf{K}_{v,n}^{-1} \end{bmatrix}\geq 0\\
\end{split}
\end{equation*}

\indent By taking simple replacements on the original formula, the proof is complete.
\end{proof}
\vspace{0.5cm}

\textit{B. Proof of Theorem \ref{thm_04_01}}\\
\indent Before showing the proof of the theorem, we need the following lemma.
\begin{lemma}
Consider the coding scheme as shown in Fig.\ref{CP_scheme},
\begin{equation*}
I(S^n;Y^n|V^n)=I(X^n\rightarrow Y^n|V^n)
\end{equation*}
\label{lemma04_01}
\end{lemma}

\begin{proof}
\begin{equation*}
\begin{split}
&I(S^n;Y^n|V^n)\\
=&h(Y^n|V^n)-h(Y^n|S^n,V^n)\\
=&\sum_{i=1}^n h(Y_i|Y^{i-1},V^n)-h(Y_i|Y^{i-1},S^n,V^n)\\
\stackrel{(a)}{=}&\sum_{i=1}^n h(Y_i|Y^{i-1},V^n)-h(Y_i|Y^{i-1},S^n,V^n,X^i)\\
\stackrel{(b)}{=}&\sum_{i=1}^n h(Y_i|Y^{i-1},V^n)-h(Y_i|Y^{i-1},X^i,V^n)\\
=&\sum_{i=1}^n I(X^i;Y_i|Y^{i-1},V^n)\\
=&I(X^n\rightarrow Y^n|V^n)\\
\end{split}
\end{equation*}
(a) follows from the fact that $X^i$ can be determined by $S^i$ and the outputs of the feedback link (i.e. $Y^{i-1}+V^{i-1}$). (b) follows from the Markov chain $S^n-(Y^{i-1},X^i,V^n)- Y_i$.
\end{proof}
\vspace{0.5cm}
\indent Now, we are ready to give the proof of the theorem.
\begin{proof}(sketch)
Define ${\tilde{C}}_{fb}^{noise}$ as formula (\ref{capacity_upperbound}). By the Szeg$\ddot{o}$-Kolmogorov-Krein theorem, we have
\begin{equation*}
{\tilde{C}}_{fb}^{noise}=\sup_{\lbrace X_i\rbrace-\text{stationary}} h(\mathcal{Y}|\mathcal{V})-h(\mathcal{W})
\end{equation*}
where the supremum is taken over all stationary Gaussian process $\lbrace X_i\rbrace_{i=0}^{\infty}$ of the form
$X_i=S_i+\sum_{k=1}^i b_k(W_{i-k}+V_{i-k})$ where $\lbrace S_i\rbrace_{i=0}^{\infty}$ is stationary and independent of ($\lbrace W_i\rbrace_{i=0}^{\infty}$,$\lbrace V_i\rbrace_{i=0}^{\infty}$) such that $\mathbb{E}[X_i^2]\leq P$.\\
\indent We first show that
\begin{equation}
\bar{C}_{fb,n}^{noise}\leq {\tilde{C}}_{fb}^{noise}
\label{apendix_equ_01}
\end{equation}
for all $n$. Fix $n$ and assume $(\mathbf{K}_{s,n}^{*},\mathbf{B}_n^{*})$ achieves $\bar{C}_{fb,n}^{noise}$. Consider a block-wise white process $\lbrace S_i\rbrace_{i=kn+1}^{(k+1)n}$, $0\leq k <\infty$, independent and identically distributed according to $\mathbb{N}_n(0,\mathbf{K}_{s,n}^{*})$.
\begin{equation*}
\begin{split}
&2n\bar{C}_{fb,n}^{noise}\\
=&I(X_1^n\rightarrow Y_1^n|V_{1}^n)+I(X_{n+1}^{2n}\rightarrow Y_{n+1}^{2n}|V_{n+1}^{2n})\\
\stackrel{(a)}{=}&I(S_1^n; Y_1^n|V_{1}^n)+I(S_{n+1}^{2n}; Y_{n+1}^{2n}|V_{n+1}^{2n})\\
=&h(S_1^n|V_{1}^n)+h(S_{n+1}^{2n}|V_{n+1}^{2n})-h(S_1^n|Y_1^n,V_{1}^n)-h(S_{n+1}^{2n}|Y_{n+1}^{2n},V_{n+1}^{2n})\\
=&h(S_{1}^{2n}|V_{1}^{2n})-h(S_1^n|Y_1^n,V_{1}^n)-h(S_{n+1}^{2n}|Y_{n+1}^{2n},V_{n+1}^{2n})\\
\leq &h(S_{1}^{2n}|V_{1}^{2n})-h(S_{1}^{2n}|Y_{1}^{2n},V_{1}^{2n})\\
=&I(S_{1}^{2n}; Y_{1}^{2n}|V_{1}^{2n})\\
\stackrel{(b)}{=}&I(X_{1}^{2n}\rightarrow Y_{1}^{2n}|V_{1}^{2n})\\
\stackrel{(c)}{=}&h(Y_{1}^{2n}|V_{1}^{2n})-h(W_1^{2n})\\
\end{split}
\end{equation*}
where (a) and (b) follows from Lemma \ref{lemma04_01}. (c) follows from the proof of Theorem \ref{thm03_02}. By repeating the same argument, we have
\begin{equation*}
\bar{C}_{fb,n}^{noise}\leq \frac{1}{kn}(h(Y_{1}^{kn}|V_{1}^{kn})-h(W_1^{kn}))
\end{equation*}
for all $k$. Next, we use the same technical skill as \cite{Kim10} to show the inequality (\ref{apendix_equ_01}). Define the time-shifted process $\lbrace X_i(t)\rbrace_{i=-\infty}^{\infty}$ where $X_i(t)=X_{i+t}$. Similarly define $\lbrace Y_i(t)\rbrace_{i=-\infty}^{\infty}$, $\lbrace W_i(t)\rbrace_{i=-\infty}^{\infty}$ and $\lbrace V_i(t)\rbrace_{i=-\infty}^{\infty}$. Introduce a random variable $T$, uniformly distributed over $\lbrace1,2,3,\cdots,n\rbrace$ and independent of everything else. Then it is easy to check that $\lbrace X_i(T),Y_i(T),W_i(T),V_i(T)\rbrace_{i=-\infty}^{\infty}$ is jointly stationary. Next, we define $\lbrace \tilde{X}_i,\tilde{Y}_i,\tilde{W}_i,\tilde{V}_i\rbrace_{i=-\infty}^{\infty}$ as a jointly Gaussian process with the same mean and autocorrelation as the stationary process $\lbrace X_i(T),Y_i(T),W_i(T),V_i(T)\rbrace_{i=-\infty}^{\infty}$. Thus,
\begin{equation*}
\begin{split}
&\bar{C}_{fb,n}^{noise}\\
\leq & \frac{1}{kn}(h(Y_1^{kn}(T)|V_1^{kn}(T),T)-h(W_1^{kn}(T)|T))\\
\stackrel{(a)}{=}& \frac{1}{kn}(h(Y_1^{kn}(T)|V_1^{kn},T)-h(W_1^{kn}))\\
\leq & \frac{1}{kn}(h(Y_1^{kn}(T)|V_1^{kn})-h(W_1^{kn}))\\
= & \frac{1}{kn}(h(\tilde{Y}_1^{kn}|V_1^{kn})-h(W_1^{kn}))\\
\end{split}
\end{equation*}
where $(a)$ follows from the stationarity assumption on noises $V$ and $W$. Taking $k\rightarrow \infty$, we obtain
\begin{equation*}
\bar{C}_{fb,n}^{noise}\leq h(\tilde{\mathcal{Y}}|\mathcal{V})-h(\mathcal{W})\leq {\tilde{C}}_{fb}^{noise}
\end{equation*}
\indent We now show the main idea of proving the other direction. Given $\epsilon>0$, we let $\lbrace \tilde{X}_i \rbrace_{i=-\infty}^{\infty}$ achieve ${\tilde{C}}_{fb}^{noise}-\epsilon$. Define the corresponding channel outputs as $\lbrace \tilde{Y}_i \rbrace_{i=-\infty}^{\infty}$. Then,
\begin{equation*}
\begin{split}
&\liminf_{n\rightarrow\infty}\bar{C}_{fb,n}^{noise}\\
=&\liminf_{n\rightarrow\infty}\max_{\lbrace X_i \rbrace_{i=1}^{n}} h(Y_1^n|V_1^n)-h(W_1^n)\\
\geq &\liminf_{n\rightarrow\infty} ( h(\tilde{Y}_1^n|V_1^n)-h(W_1^n))\\
=&\lim_{n\rightarrow\infty} h(\tilde{Y}_1^n|V_1^n)-h(W_1^n)\\
=&h(\tilde{\mathcal{Y}}|\mathcal{V})-h(\mathcal{W})\\
=&{\tilde{C}}_{fb}^{noise}-\epsilon\\
\end{split}
\end{equation*}
Taking $\epsilon\rightarrow 0$, we obtain
\begin{equation*}
\liminf_{n\rightarrow\infty}\bar{C}_{fb,n}^{noise}\geq {\tilde{C}}_{fb}^{noise}
\end{equation*}
The technical discussion on power constraint is identical to that in \cite{Kim10}, so we herein omit it. Combined with inequality (\ref{apendix_equ_01}), we know that the limit of $\bar{C}_{fb,n}^{noise}$ exists and
\begin{equation*}
\lim_{n\rightarrow\infty}\bar{C}_{fb,n}^{noise}={\tilde{C}}_{fb}^{noise}.
\end{equation*}
\end{proof}

%% file: Allerton11.bbl
\begin{thebibliography}{10}
\providecommand{\url}[1]{#1}
\csname url@samestyle\endcsname
\providecommand{\newblock}{\relax}
\providecommand{\bibinfo}[2]{#2}
\providecommand{\BIBentrySTDinterwordspacing}{\spaceskip=0pt\relax}
\providecommand{\BIBentryALTinterwordstretchfactor}{4}
\providecommand{\BIBentryALTinterwordspacing}{\spaceskip=\fontdimen2\font plus
\BIBentryALTinterwordstretchfactor\fontdimen3\font minus
  \fontdimen4\font\relax}
\providecommand{\BIBforeignlanguage}[2]{{%
\expandafter\ifx\csname l@#1\endcsname\relax
\typeout{** WARNING: IEEEtran.bst: No hyphenation pattern has been}%
\typeout{** loaded for the language `#1'. Using the pattern for}%
\typeout{** the default language instead.}%
\else
\language=\csname l@#1\endcsname
\fi
#2}}
\providecommand{\BIBdecl}{\relax}
\BIBdecl

\bibitem{cover89}
T.~M. Cover and S.~Pombra, ``Gaussian feedback capacity,'' \emph{IEEE
  Transactions on Information Theory}, vol.~35, no.~1, pp. 37--43, 1989.

\bibitem{shannon49}
C.~E. Shannon, ``Communication in the presence of noise,'' \emph{Proc.IRE},
  vol.~37, pp. 10--21, 1949.

\bibitem{Kim10}
Y.~H. Kim, ``Feedback capacity of stationary gaussian channels,'' \emph{IEEE
  Transactions on Information Theory}, vol.~56, no.~1, pp. 57--85, 2010.

\bibitem{Chance10}
Z.~Chance and D.~J. Love, ``A noisy feedback encoding scheme for the gaussian
  channel,'' \emph{IEEE International Conference on Acoustics Speech and Signal
  Processing}, pp. 3482--3485, 2010.

\bibitem{chong11_ISIT}
C.~Li and N.~Elia, ``Bounds on the achievable rate of noisy feedback gaussian
  channels under linear feedback coding scheme,'' \emph{IEEE International
  Symposium on Information Theory}, pp. 71--75, 2011.

\bibitem{Kim07}
Y.~H. Kim, A.~Lapidoth, and T.~Weissman, ``The gaussian channel with noisy
  feedback,'' \emph{IEEE International Symposium on Information Theory}, pp.
  1416--1420, 2007.

\bibitem{Wyner69}
A.~D. Wyner, ``On digital communication over a discere-time gaussian channel
  with noisy feedback,'' \emph{The Bell System Technical Journal}, pp.
  3173--3186, Dec.1969.

\bibitem{Martins08}
N.~C. Martins and T.~Weissman, ``Coding for additive white noise channels with
  feedback corupted by quantization or bounded noise,'' \emph{IEEE Transactions
  on Information Theory}, vol.~54, no.~9, pp. 4274--4282, Sep.2008.

\bibitem{chance11}
Z.~Chance and D.~J. Love, ``Concatenated coding for the awgn channel with noisy
  feedback,'' \emph{Arxiv preprint arXiv:0909.0105}, 2011.

\bibitem{Schalkwijk66}
J.~P.~M. Schalkwijk and T.~Kailath, ``A coding scheme for additive noise
  channels with feedback i: No bandwidth constraint,'' \emph{IEEE Transactions
  on Information Theory}, vol. IT-12, no.~2, pp. 172--182, 1966.

\bibitem{cover06}
T.~M. Cover and J.~A. Thomas, \emph{Elements of Information Theory, 2nd
  edition}.\hskip 1em plus 0.5em minus 0.4em\relax New York: Wiley, 2006.

\bibitem{Massey1990}
J.~L. Massey, ``Causality, feedback and directed information,'' \emph{Proc. of
  the IEEE Conf on Decision and Control}, 2002.

\bibitem{Tati09}
S.~Tatikonda and S.~Mitter, ``The capacity of channels with feedback,''
  \emph{IEEE Transactions on Information Theory}, vol.~55, no.~1, pp. 323--349,
  2009.

\bibitem{ConvexOpti}
S.~Boyd and L.~Vandenberghe, \emph{Convex Optimization}.\hskip 1em plus 0.5em
  minus 0.4em\relax Cambridge: Cambridge university, 2004.

\bibitem{chong_jour_nf}
C.~Li and N.~Elia, ``The information flow and capacity of channels with noisy
  feedback,''
  \emph{http:$//arxiv.org/PS_cache/arxiv/pdf/1108/1108.2815v1.pdf$}, 2011.

\end{thebibliography}
